\documentclass[prl,twocolumn,showpacs,amsmath,amssymb,superscriptaddress]{revtex4-1}
\usepackage{dcolumn}
\usepackage{bm,graphicx}
\usepackage{etoolbox}
\usepackage{url}
\usepackage[colorlinks]{hyperref}
\usepackage{breakurl}
\usepackage{color}
\usepackage{amsmath}
\usepackage{xcolor}
\usepackage{soul}

\newcommand{\ee}{\varepsilon}
\newcommand{\kk}{\mathbf{k}}

\newcommand{\s}{\sigma}

\newcommand{\om}{\omega}
\newcommand{\al}{\alpha}
\newcommand{\D}{\Delta}

\def\mathclap#1{\text{\hbox to 0pt{\hss$\mathsurround=0pt#1$\hss}}}

\begin{document}

\title{Two-dimensional conical dispersion in ZrTe$_5$ evidenced by optical spectroscopy}

\author{E.~Martino}\thanks{These authors contributed equally.}
\affiliation{IPHYS, EPFL, CH-1015 Lausanne, Switzerland}

\author{I.~Crassee}\thanks{These authors contributed equally.}
\affiliation{LNCMI, CNRS-UGA-UPS-INSA, 25, avenue des Martyrs, F-38042 Grenoble, France}

\author{G.~Eguchi}
\affiliation{Institute of Solid State Physics, Vienna University of Technology, Wiedner Hauptstrasse 8-10, 1040 Vienna, Austria}

\author{D. Santos-Cottin}
\affiliation{Department of Physics, University of Fribourg, Chemin du Mus\'ee 3, CH-1700 Fribourg, Switzerland}

\author{R.D.~Zhong}
\author{G.D.~Gu}
\affiliation{Condensed Matter Physics and
   Materials Science Department, Brookhaven National Laboratory, Upton,
   New York 11973, USA}

\author{H.~Berger}
\affiliation{IPHYS, EPFL, CH-1015 Lausanne, Switzerland}

\author{Z.~Rukelj}
\affiliation{Department of Physics, Faculty of Science, University of Zagreb, Bijeni\v{c}ka 32, HR-10000 Zagreb, Croatia}

\author{M.~Orlita}
\affiliation{LNCMI, CNRS-UGA-UPS-INSA-EMFL, 25, avenue des Martyrs, F-38042 Grenoble, France}
\affiliation{Institute of Physics, Charles University in Prague, CZ-12116 Prague, Czech Republic}

\author{C.~C.~Homes}
\affiliation{Condensed Matter Physics and
   Materials Science Department, Brookhaven National Laboratory, Upton,
   New York 11973, USA}

\author{Ana Akrap}\email{ana.akrap@unifr.ch}
\affiliation{Department of Physics, University of Fribourg, Chemin du Mus\'ee 3, CH-1700 Fribourg, Switzerland}

\date{\today}

\begin{abstract}
Zirconium pentatelluride was recently reported to be a 3D Dirac semimetal, with a single conical band, located at the center of the Brillouin zone. The cone's lack of protection by the lattice symmetry immediately sparked vast discussions about the size and topological/trivial nature of a possible gap opening. Here we report on a combined optical and transport study of ZrTe$_5$, which reveals an alternative view of electronic bands in this material. We conclude that the dispersion is approximately linear only in the $a$-$c$ plane, while remaining relatively flat and parabolic in the third direction (along the $b$ axis). 
Therefore, the electronic states in ZrTe$_5$  cannot be described using the model of 3D Dirac massless electrons, even when staying at energies well above the band gap $2\D = 6$ meV found in our experiments at low temperatures. 
\end{abstract}

\maketitle

In materials with topological phases, small energy scales can play an important role. ZrTe$_5$  is an excellent example. The band gap opening at the center of the Brillouin zone is caused by a strong spin orbit interaction \cite{Weng2014}, making the gap topological, be it positive, zero, or negative. ZrTe$_5$ is a layered compound with an extremely high mobility, and there is consensus in scientific literature that the low energy bands in ZrTe$_5$ are conical \cite{Li2016,Manzoni2015,Xiong2017,Chen2015,Chen2015m,Liu2016}. However, the delicate balance of these energy scales has led to many contradicting reports.
Several possible topological phases were predicted or reported in ZrTe$_5$, amongst them a quantum spin Hall insulator \cite{Weng2014}, weak topological insulator (WTI) \cite{Moreschini2016}, strong TI (STI) \cite{Manzoni2016,Chen2017, Xu2018}, and a three-dimensional (3D) Dirac semimetal \cite{Chen2015,Chen2015m}. All of these possible phases are linked to a key question: What is the true dimensionality of the conical dispersion in ZrTe$_5$? The detailed band structure has not yet been established, nor is it known whether the linear dispersion is indeed three-dimensional. Band structure calculations critically depend upon fine structural details \cite{Fan2017}. ARPES measurements have shown linearly dispersing bands in the $a$-$c$ planes, and a strongly varying chemical potential as a function of temperature \cite{Manzoni2015,Xiong2017}. It is an open question how the shift of chemical potential measured at the surface relates to the bulk properties, and what the dispersion in the out-of-plane direction is.

In this work, we demonstrate a two-dimensional conical dispersion, and show the temperature-induced shift of the chemical potential across the gap in ZrTe$_5$. 
Our findings are based on bulk-sensitive techniques, optical spectroscopy and magneto-transport. We address low-energy states due to low carrier density in our samples. We show that the free-carrier optical plasmon energy depends non-monotonically on temperature. The sign of the dominant carriers changes from high-temperature thermally-activated holes to low-temperature electrons. 
Most importantly, we find that the energy dispersion cannot be linear in all three directions. Rather, our optical conductivity points to a linear dispersion in the $a$-$c$ plane, and a parabolic dispersion along the $b$ axis. We construct an effective Hamiltonian explaining both the optical and transport properties at low temperatures. 
Our results place a strong doubt over the commonly accepted picture of a 3D Dirac dispersion.

\begin{figure}[h]
    \includegraphics[trim = 0mm 0mm 0mm 0mm, clip=true, width=\linewidth]{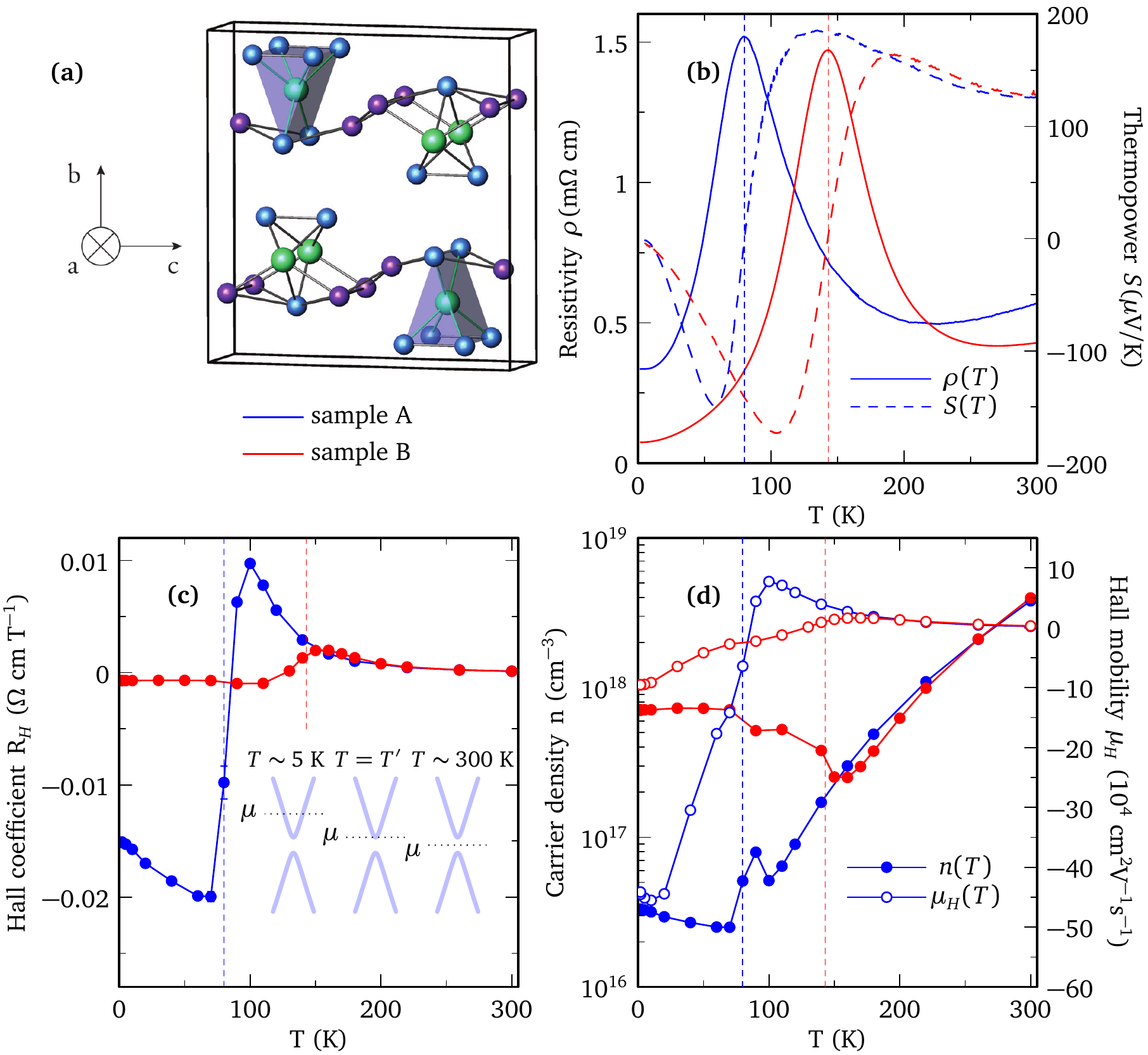}
\caption{\label{Fig1:Transport} (color online) (a) Orthorhombic unit cell of ZrTe$_5$. (b) Resistivity and thermoelectric power, (c) Hall coefficient, (d) carrier density  and Hall mobility are shown for samples A and B. Blue and red dashed vertical lines denote temperatures $T'_A$ and $T'_B$, respectively. Inset in (c) illustrates the chemical potential shift with temperature.}
\end{figure}

Measurements were performed on samples synthesized by two different methods, self-flux growth \cite{Li2016} (sample A) and chemical vapor transport \cite{Levy1983} (sample B), leading to different low-temperature carrier concentrations. The transport properties are measured using a custom setup. The magneto-transport data are obtained using Quantum Design PPMS. Optical reflectance is measured using FTIR spectroscopy, with {\em in situ} gold evaporation \cite{Homes1993}. At high energies, the phase was fixed by ellipsometry. We use Kramers-Kronig relations to obtain the frequency-dependent complex dielectric function $\epsilon(\omega)$, where $\omega$ is the incident photon frequency. 
Magneto-transmission was measured using a superconducting coil, with sample at $T=2$~K in a low-pressure helium exchange gas. Analysis of the optical spectra was performed using RefFIT software \cite{Kuzmenko2005}.

The orthorhombic structure of ZrTe$_5$ is shown in Fig.~\ref{Fig1:Transport}(a). The most conducting direction is the $a$ axis, running along the Zr chains. The layers are stacked along the least conducting $b$ direction. The conduction and valence bands are based upon the tellurium $p$ orbitals.
Figure \ref{Fig1:Transport}(b--d) shows electronic transport along the $a$ axis for samples A and B. 
Panel (b) shows resistivity $\rho$ and thermoelectric power $S$, (d) Hall coefficient $R_H$, (e) single band carrier density $n$, and Hall mobility $\mu_H$, each as a function of temperature. A dramatic change occurs in each quantity at temperature $T'$; $T'_A=80$~K for sample A, and $T'_B=145$~K for sample B. These temperatures correspond to a maximum in $\rho$, a sign inversion in $S$, $R_H$ and $\mu_H$, and a minimum in $n$.
The resistivity peak appears to be linked to a minimum in carrier density at $T'$, with a concomitant crossover from electron to hole-dominated conduction. 

The metallic resistivity well below $T'$ is described by $\rho=\rho_0+AT^2$, with $A_A=0.1~\mu\Omega$cm/K$^{2}$ and $A_B=0.036~\mu\Omega$cm/K$^{2}$ for sample A and sample B, respectively. The coefficient $A$ is inversely proportional to $E_F$ \cite{Lin2015}, indicating that the Fermi level in sample A is lower than in sample B.
The Mott formula $S(T)=k_B^2 T/(e E_F)$ gives an estimate of the low-temperature Fermi levels for samples A and B, $E_F^A\sim 14$~meV and $E_F^B \sim 23$~meV. 
The lower Fermi level in sample A is consistent with a lower carrier density (Fig.~\ref{Fig1:Transport}(e)).
The Hall coefficient, carrier density, and Hall mobility are obtained in a single band analysis in the $B\rightarrow 0$ limit. They strongly differ for the two samples below $T'$. In sample A, the mobility at 2~K is extremely high: $\mu_H^A = 0.45\times 10^{6}$ cm$^2/$(Vs), whereas the carrier density is $n_A = 3\times10^{16}$cm$^{-3}$, surprisingly low for a metallic system. 
A two-band model \cite{SM, Eguchi2019} shows that minority carriers contribute very little to low-temperature conductivity. However, close to $T'$ a two-band picture is needed.

Above 180 K, $S$, $R_H$, $n$ and $\mu_H$ are similar in both samples, suggesting that the thermally activated carriers dominate at high temperatures. 
At room temperature, both samples show weakly metallic resistivity, while thermopower is activated, $S=C+ 2\Delta/(eT)$, giving a band gap of $2\Delta \sim 20$ meV at high temperature for both samples ($C$ is a constant offset). 
The chemical potential is therefore within the gap at high temperature. 
Its temperature evolution is illustrated by the inset in Fig.~\ref{Fig1:Transport}(c); $T'$ depends on the low-temperature carrier density. A small band gap and a steep band dispersion may lead to a strong shift of chemical potential, consistent with linear dispersion.

\begin{figure}[t]
\includegraphics[trim = 0mm 0mm 0mm 0mm, clip=true, width=1\linewidth]{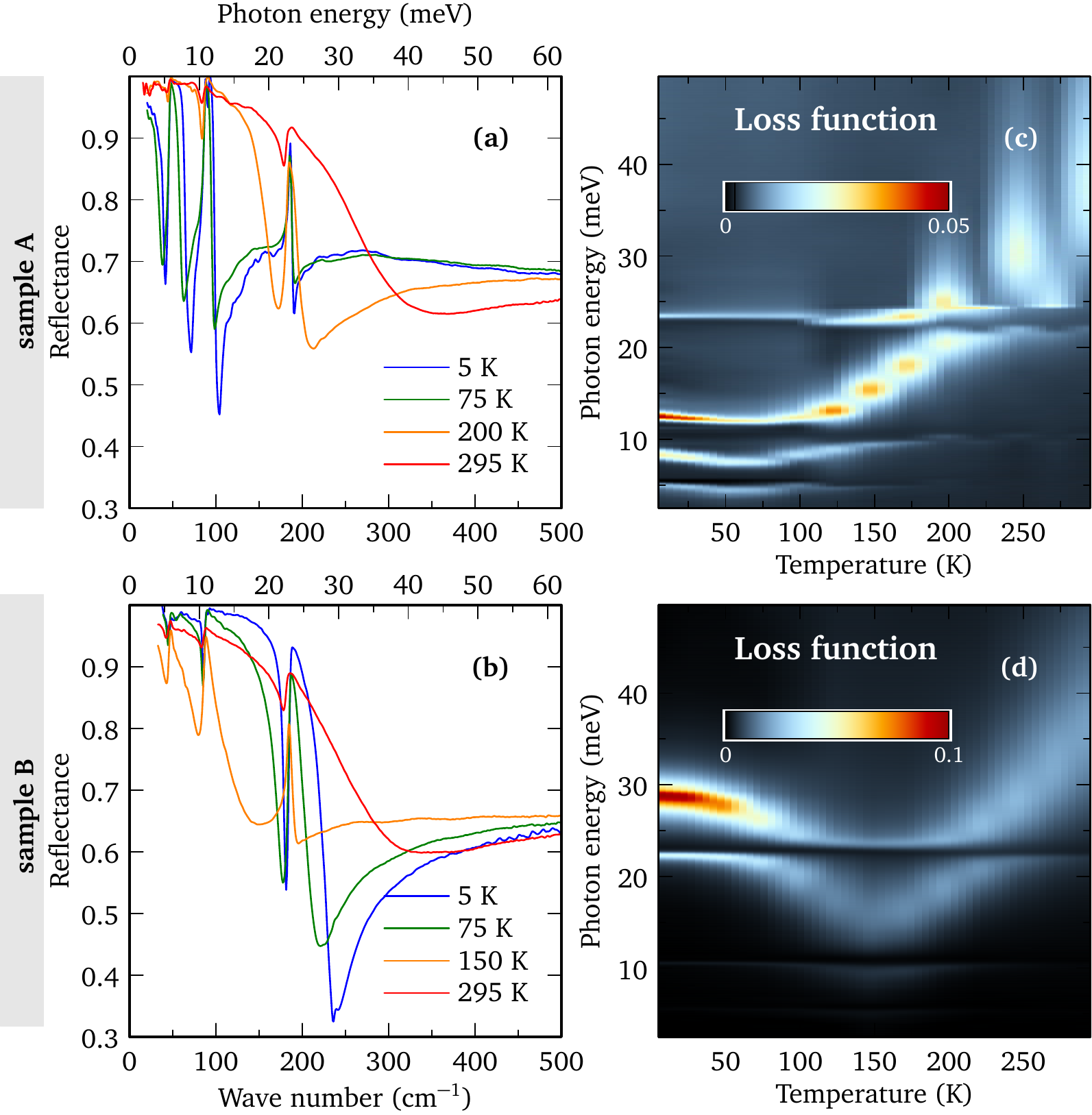}
\caption{\label{Fig2:ColorPlots} (color online) Reflectance of (a) sample A and (b) sample B as a function of photon energy. Loss function colormap for (b) sample A and (d) sample B. The data was taken at each 25K, and interpolated.}
\end{figure}

\begin{figure*}[t]
\includegraphics[trim = 0mm 0mm 0mm 0mm, clip=true, width=1\linewidth]{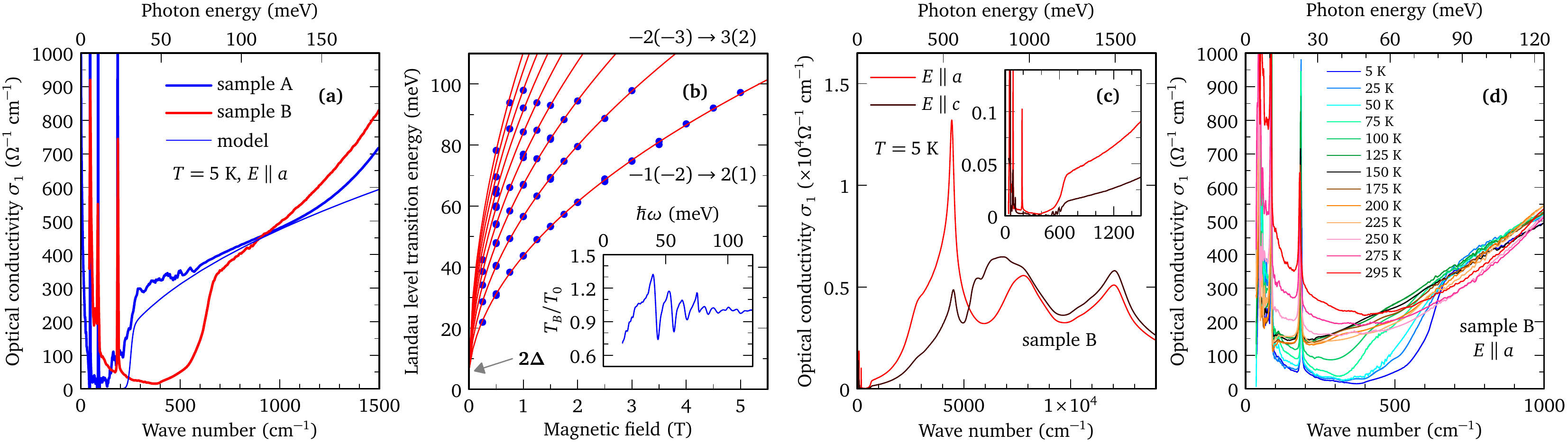}
\caption{\label{Fig3:Optics} (color online)  (a) $\sigma_1(\omega)$ is shown at 5 K for $E \parallel a$,  the calculated conductivity ($\sim \sqrt{\omega}$) is detailed in the text. (b) Landau level transition energies obtained from magneto-optical transmission measurements on sample A, at $T=2$~K. Red lines are a fit to Eq.~\ref{LLs}; the gap $2\Delta=6$~meV is indicated. The two lowest observed transitions are labeled. Inset shows a relative magneto-transmission spectrum for $B=1$~T.
(c)  Wide frequency range $\sigma_1(\omega)$ for the two polarizations in the $a$-$c$ plane, at $T=5$~K; inset shows low-energy details of $\sigma_1$. (d) Temperature evolution of $\sigma_1(\omega)$ for $E \parallel a$. }
\end{figure*}

We have identified the maximum in the resistivity with the minimum in the carrier density. However, resistivity also depends on the scattering rate. 
To show that it is the carrier density, and not the scattering mechanism, which dominantly drives the resistivity maximum, one can determine the optical properties. 
The reflectance plasma edge $\omega_p$ is linked to the carrier concentration, $\omega_p^2 \propto n/m$, and does not depend on the scattering rate.

The reflectance and loss function are shown in Fig.~\ref{Fig2:ColorPlots}
High reflectance at low energies ($R \rightarrow 1$ as $\omega \rightarrow 0$) is followed by a sharp drop at the plasma edge. The room-temperature reflectance looks very similar in samples
A and B, with a broad plasma minimum at 45~meV, confirming that the carriers are thermally activated at high temperatures.
At low temperatures, the plasma edge in sample A is lower and the phonon-related features are much more pronounced,
consistent with a lower carrier density and poorer screening.
The plasma edge is lower at 75~K than at 5~K, signifying there is a loss of itinerant carriers as 
temperature increases from 5 to 75~K. Similarly, in sample B, the plasma edge is the lowest for $T\simeq150$~K.

The non-monotonic change in carrier density can be visualized better by plotting the loss function, defined as $-\rm{Im} (1/\tilde\epsilon)$, showing the collective modes as peaks. 
The itinerant carrier plasmon appears as a strong peak with a non-monotonic temperature dependence (Fig.~\ref{Fig2:ColorPlots}(c-d)), with minima at $T'_A$ and $T'_B$ for samples A and B.
The loss function also shows three phonon plasmons, visible as horizontal lines in Fig.~\ref{Fig2:ColorPlots}(c).
Overall, the reflectance and loss function firmly establish that the carrier density changes non-monotonically with temperature in bulk ZrTe$_5$, 
in full agreement with the temperature dependence of the carrier density in Fig.~\ref{Fig1:Transport}(e). The peak in $\rho(T)$ is therefore intimately
linked to the local minimum in $n(T)$. The same effect is seen in samples with more than an order of magnitude difference in carrier density.

The optical conductivity is related to the dielectric function by $\s(\omega)=-2\pi i \omega [{\epsilon}(\omega)-\epsilon_\infty]/Z_0 =\sigma_1(\omega)+ i \sigma_2(\omega),$
where $Z_0\approx 377\, \Omega$ is the impedance of free space. 
The real part of optical conductivity, $\sigma_1(\omega)$, is shown in Fig.~\ref{Fig3:Optics}(a) 
for samples A and B at 5 K, for light polarized along the $a$ axis. The optical conductivity is dominated by a sharp 
Pauli blocking edge; interband transitions are allowed only when the incident photon energy is higher than the optical gap $2\ee_F$, with Fermi level $\ee_F$  measured from the band gap middle. 
The lower onset of interband absorption in sample A, than in sample B, is consistent with the lower 
$\ee_F$ in sample A. The optical gap is $2\ee_F = 28$ meV in sample A, and 74 meV in sample B.
The band gap was determined using magneto-optical transmission measurements, giving $2\D = 6$ meV. As discussed below, this band gap is deduced from a small, but well-defined deviation of the interband inter-Landau level transitions from a $\sqrt{B}$ dependence, which is otherwise typical of massless (gapless) charge carriers. 
This points to a linear dispersion in the $a$-$c$ plane, in agreement with previous work \cite{Chen2015m,Chen2017,Jiang2017}. 

A combined analysis of the optical conductivity and Hall effect data allows us to determine the effective cyclotron masses in both samples. Spectral weight analysis is performed by integrating the Drude part of the conductivity curve up to $\omega^* = 5$~meV, such that at 5~K the Drude contribution falls within the integration limit \cite{SM}. 
The effective $a$-$c$ plane masses for samples A and B obtained from this analysis are shown in Table \ref{tab:m_eff}. Importantly, the effective mass of sample A is smaller than for sample B: $m_B \approx 2.5 m_A$. The apparent dependence of the effective mass on the Fermi energy is clear evidence for a non-parabolic dispersion. For a linearly dispersing system, the effective (cyclotron) mass may be defined by $\ee_F/v_a^2$, which is in excellent agreement with $m$ (Table \ref{tab:m_eff}), and indicates conical dispersion in the $a$-$c$ plane.
\begin{table}[tb]
\caption{Hall mobility, Hall carrier density, effective (cyclotron) mass, optical gap $2\ee_F$, and $\ee_F/v_a^2$ at 5 K. 
Velocity along $a$ axis is $v_a=6.9\times 10^5$ m/s.}
\vspace*{0.1cm}
\begin{ruledtabular}
\begin{tabular}{c||ccccc}
sample &$\mu_H$ (cm$^{2}/$(Vs)) &  $n$(cm$^{-3}$) & $m$  & $2\ee_F$ (meV) & $\ee_F/v_a^2$ \\
 \cline{1-6}
A &$4.5\times10^{5}$ &  $3\times 10^{16}$ & 0.0052  & 28 & 0.0052 \\
B & $1.0\times 10^{5}$ &  $7\times 10^{17}$ & 0.0125  & 74 & 0.0137 \\
\end{tabular}
\end{ruledtabular}
\label{tab:m_eff}
\end{table}

We now want to verify the nature of the linear dispersion. For a 3D conical band, one expects the real part of optical conductivity to grow linearly with the frequency, $\s_1(\omega) = {e^2  \nu \omega }/({12 h v_F}) $, where $\nu$ is the number of nondegenerate cones at the Fermi level \cite{Hosur2012,Bacsi2013}.
Such dependence is indeed observed for sample B, with a higher optical gap. However, this model fatally fails to explain the optical conductivity of sample A (Fig.~\ref{Fig3:Optics}(a)) with a lower doping, where  $\s_1(\omega)$ increases quasi-linearly with $\omega$, but with a well-defined offset. 

It has been proposed that such an offset may arise from self-energy effects, $\sigma_1(\omega) \propto \omega-4\Delta$, which may induce a positive or negative band gap $2\Delta$ \cite{Tabert2016, Neubauer2016}. Adopting this scenario, our data would imply $2\Delta\sim -50$~meV (Fig.~\ref{Fig3:Optics}(a)). However, this value exceeds, by an order of magnitude, the size of the gap directly measured by magneto-transmission experiments (Fig.~\ref{Fig3:Optics}(b)). Moreover, the gap readout from the magneto-optical data -- if indeed due to self-energy effects -- would have to be positive.

To explain the linear, but clearly offset optical conductivity, we propose a simple effective Hamiltonian. It differs from the 3D massive Dirac electron model, often used for ZrTe$_5$, but still implies a massive Dirac dispersion in the $a$-$c$ plane, with a parabolic dispersion around the band gap $2\Delta$ that straightens to a linear dispersion at higher energies. The dispersion along the $b$ direction remains parabolic or Schr\"{o}dinger-like at all relevant energies:
\begin{equation}\label{jed}
H=\begin{pmatrix}
 \Delta + \zeta k_b^2 & \hbar v_ak_a - i \hbar v_ck_c \\
 \hbar v_ak_a + i \hbar v_ck_c &  -\Delta - \zeta k_b^2
\end{pmatrix}.
\end{equation}
$v_{\al}$ are the Dirac velocities, and $\zeta = \hbar^2/2m^*$ where $m^*$ is the $b$-direction effective mass. The eigenvalues of the Hamiltonian are 
$\ee_{2,1 \kk} = \pm \sqrt{\hbar^2(v_ak_a)^2 + \hbar^2(v_ck_c)^2 + (\Delta + \zeta k_b^2 )^2 }$
and they are symmetrical with respect to the band gap middle.
The interband conductivity along the $a$ axis can be evaluated in the relaxation constant approximation for $T \approx 0$ \cite{SM}
\begin{eqnarray}\label{dev}
 \s_1^a(\omega) =  \frac{e^2}{2\pi\hbar^2} \frac{v_a}{v_c} \sqrt{m^*}\sqrt{\hbar \omega - 2\D} \, \,\,\Theta(\hbar\omega - 2\ee_F)
\end{eqnarray}
The ratio $v_a/v_c \approx 1.5$ is determined from the ratio of interband conductivities along the $a$ and $c$ axes (Fig.~\ref{Fig3:Optics}(c)), leaving $m^*$ as the only fitting parameter. 
The fit shown in Fig.~\ref{Fig3:Optics}(a) gives $m^*\approx 1.8 m_e$, and matches the experimental optical conductivity very well, confirming that the dispersion in ZrTe$_5$ is linear in the $a$-$c$ plane, and parabolic along $b$ direction.
Based on the above Hamiltonian, we can determine the total carrier concentration \cite{SM}:
\begin{equation}\label{r2}
 n =   \frac{1}{\pi^2 \hbar^3} \frac{1}{v_av_c} \sqrt{2m^*} \frac{2}{15} (\ee_F-\D)^{3/2}(2\Delta + 3\ee_F).
\end{equation}
Using the value from the Hall effect, $n = 3\times10^{16}$~cm$^{-3}$ for sample A, we obtain $v_a = 7.0\times10^5$~m/s  and $v_c = 4.6\times10^5$~m/s, in very good agreement with Shubnikov de Haas experiments \cite{Liu2016}.
The bare plasmon energy is 
$\hbar  \om_{pl} = \hbar \sqrt{e^2 n_{aa}/(\epsilon_0 m_e)} = 0.12 \, \rm{eV} $,
in good agreement with the experimental fit for sample A, 0.1~eV  \cite{SM}.
The energy dispersion may be expanded for small values of $k_a$, $k_b$ and $k_c$, since the conduction band is weakly filled. 
The expansion gives a closed Fermi surface of ellipsoidal shape whose effective masses in various directions are $m_{a}=\Delta/v_a^2=0.001 m_e$, $m_{c}=\Delta/v_c^2=0.0025 m_e$, and $m_{b}= m^*=1.8m_e$.
The Landau levels for the Hamiltonian (\ref{jed}) for a magnetic field applied along $b$ axis are \cite{SM}
\begin{equation}\label{LLs}
 \tilde{\ee}(B) = \pm \sqrt{2\hbar e v_a v_c B \mathcal{N}  + \Delta^2 }.
\end{equation}
The fit in Fig.~\ref{Fig3:Optics}(b) gives a band gap of $2\D=6$~meV, and the effective Fermi velocity $\sqrt{v_a v_c} = 4.9\times 10^5$~m/s. 

The presence of a band gap in ZrTe$_5$ agrees with the DFT \cite{Weng2014,Fan2017,Manzoni2016,SM}, nevertheless, its size appears to be overestimated in these calculations.
The DFT favors STI over WTI as a ground state of ZrTe$_5$, both in monolayer and bulk form. Nevertheless, the DFT theory appears to overestimate its size (25-100 meV).
Experimentally, the situation is less clear. Both STI and WTI phases have been reported by ARPES or STM/STS \cite{Moreschini2016,Manzoni2016,Xiong2017,LiSTM2016,Wu2016}.
While we do not find direct evidence of either STI or WTI in our experimental data, such a conclusion was made in a recent magneto-optical study \cite{Chen2017}, reporting on crossing of zero-mode Landau levels, typical of STIs. The DFT studies also indicate \cite{Weng2014,Fan2017} that the out-of-plane dispersion is considerably flatter as compared to the in-plane one. This is in line with our findings and the layered nature of ZrTe$_5$. At higher energies, our optical spectra agree with those determined by DFT calculation \cite{SM}.

Figure~\ref{Fig3:Optics}(d) shows $\sigma_1(\omega)$ for sample B, taken at many different temperatures. 
As the temperature increases from 5~K to 150~K, the Pauli edge gradually smears out and shifts to lower energies, consistent with the decrease of carrier density. Interestingly, $\sigma_1$ appears to be linear in $\omega$ at $T=150$~K.
Above 150~K, the low frequency range is filled out by a Drude contribution of the thermally excited carriers which become accessible for $T > 2\Delta/k_B$.

The scattering rate $\gamma$ for the Drude contribution can be obtained from a Drude-Lorentz modelling of the reflectance  \cite{Kuzmenko05}. At 5K, for sample A one obtains $\gamma = 1\pm 1$~meV.
The scattering rate can also be extracted from $\sigma_{dc}=e^2n_{aa}\hbar/(m_e \gamma)$, where $n_{aa}$ is obtained from our model calculation \cite{SM}.  
Here, $n_{aa}$ is the spectral weight of the Drude contribution and is finite irregardless of temperature.
This gives $\gamma = 0.5$~meV for sample A, within the error bars of the optically determined  scattering rate.

The Hamiltonian (\ref{jed}) may also quantitatively explain the observed $T^2$ behavior in the resistivity. The $T^2$ resistivity dependence in a 3D metal is usually caused by three mechanisms: Umklapp scattering, Koshino-Taylor impurity scattering \cite{Koshino1960,Taylor1964}, and thermal activation of carriers. The latter is linked to the temperature dependence of the chemical potential, which is significant in ZrTe$_5$. The electron band properties allow us to calculate the expected coefficient for the sample A, giving two thirds of the fit to the experimental data, $A^{calc}_A \simeq 2/3  A_A$ \cite{SM}. 
All of this points to a fairly good agreement between the model, the optical results, and the transport results.

Finally, Fig.~\ref{Fig3:Optics}(c) shows the optical conductivity in a broad frequency range for both in-plane polarizations, at $T=5$~K. Several strong features are apparent, and the strongest is at 0.5~eV, which is only $\sim 50$~meV wide for $E\parallel a$. This feature is a van Hove singularity, due to transitions between flat bands, and it indicates a weaker dispersion along $b$ axis, fully consistent with our Hamiltonian.

In conclusion, ZrTe$_5$ is a fairly simple two-band system of extremes. It has a small band gap, very small effective mass, and may reach extremely low carrier concentration, yet showing metallic conductivity with very high mobility. These specific physical characteristics lead to a chemical potential that strongly shifts as a function of temperature.
Crucially, the optical conductivity clearly contradicts the scenario of a 3D cone. Based on the characteristic frequency dependence of $\sigma_1(\omega)$, we conclude that while the dispersion is linear in the $a$-$c$ plane well above the band gap, it remains parabolic along $b$ axis.

The authors acknowledge illuminating discussions with K. Behnia, B. Fauqu\'e, A.B. Kuzmenko, D. van der Marel, and A. Soluyanov, and kind help by N. Miller. 
We also thank A. Crepaldi for his generous help with samples, and for extensive discussions.
I.~C. acknowledges funding from the Postdoc.Mobility fellowship of the Swiss National Science Foundation.
A.~A. acknowledges funding from the  Swiss National Science Foundation through project PP00P2\_170544.
This work has been supported by the ANR DIRAC3D. We acknowledge the support of LNCMI-CNRS, a member of the European Magnetic Field Laboratory (EMFL).
Work at BNL was supported by the U.S. Department of Energy, Office of Basic Energy Sciences, Division of Materials Sciences and Engineering under Contract No. DE-SC0012704.

\bibliography{ZrTe5}

\newpage \ \\

\newpage
\vspace*{-2.0cm}
\hspace*{-2.5cm} {
  \centering
  \includegraphics[width=1.2\textwidth]{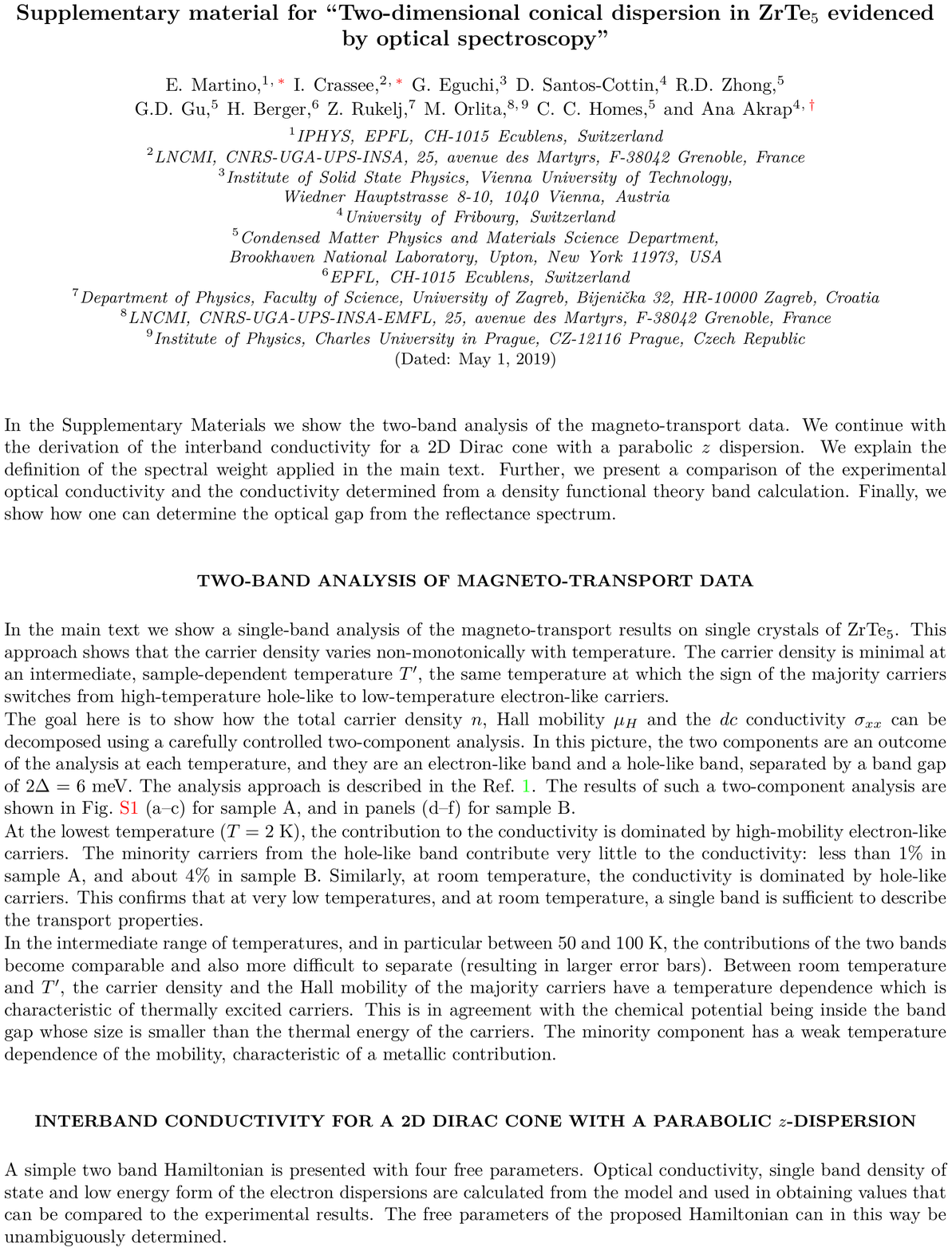} \\
  \ \\
}

\newpage
\vspace*{-2.0cm}
\hspace*{-2.5cm} {
  \centering
  \includegraphics[width=1.2\textwidth]{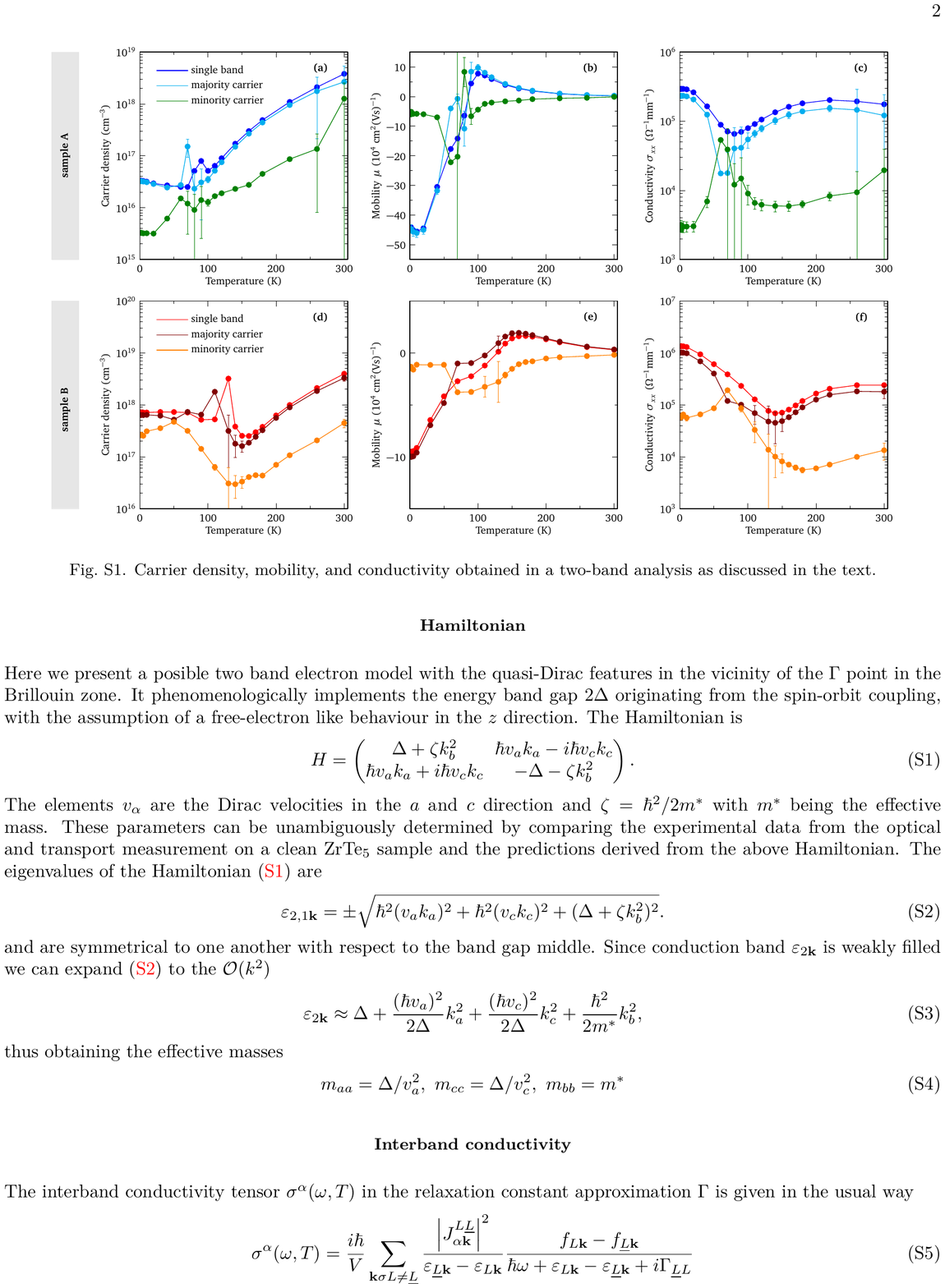} \\
  \ \\
}

\newpage
\vspace*{-2.0cm}
\hspace*{-2.5cm} {
  \centering
  \includegraphics[width=1.2\textwidth]{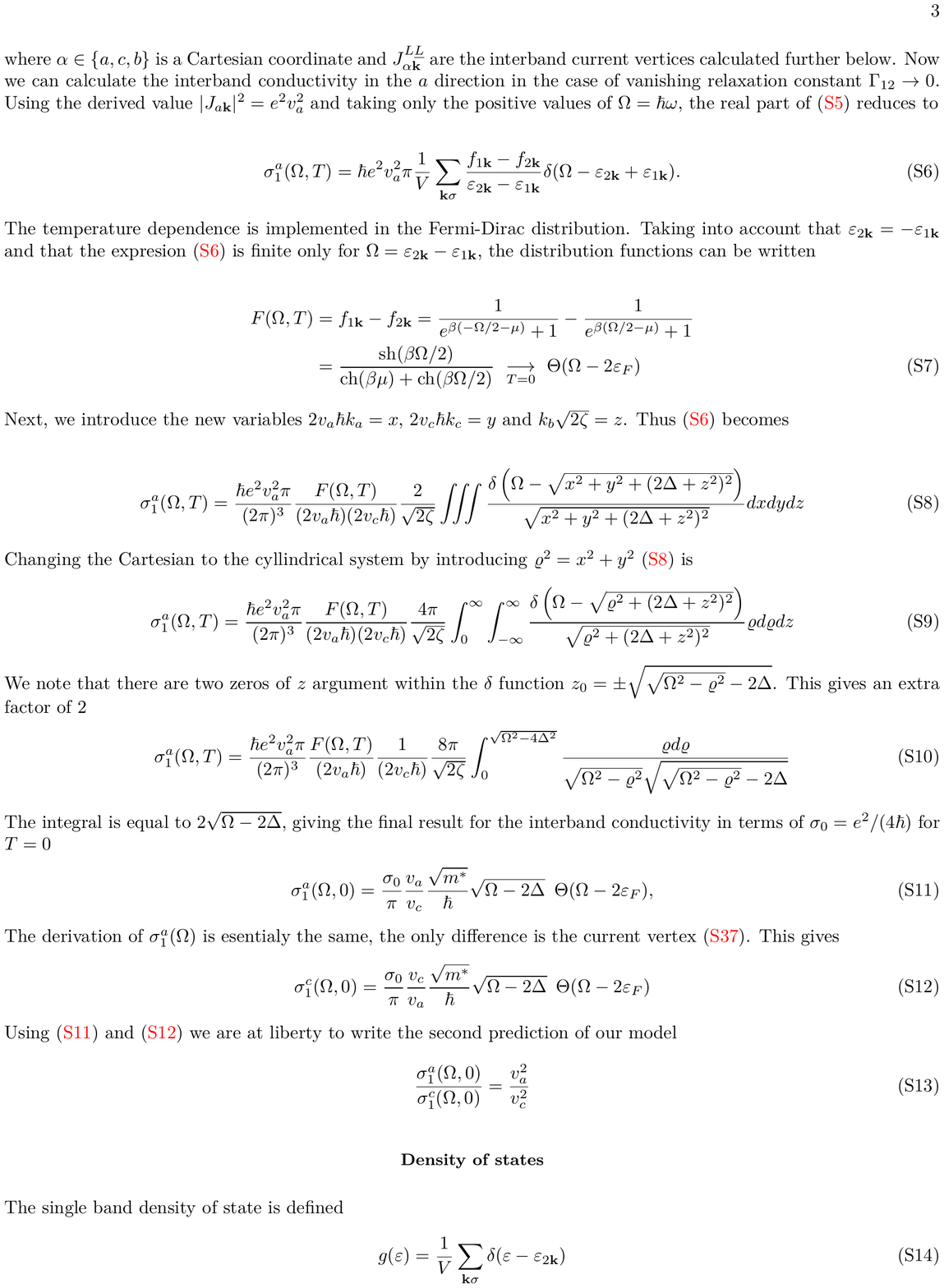} \\
  \ \\
}

\newpage
\vspace*{-2.0cm}
\hspace*{-2.5cm} {
  \centering
  \includegraphics[width=1.2\textwidth]{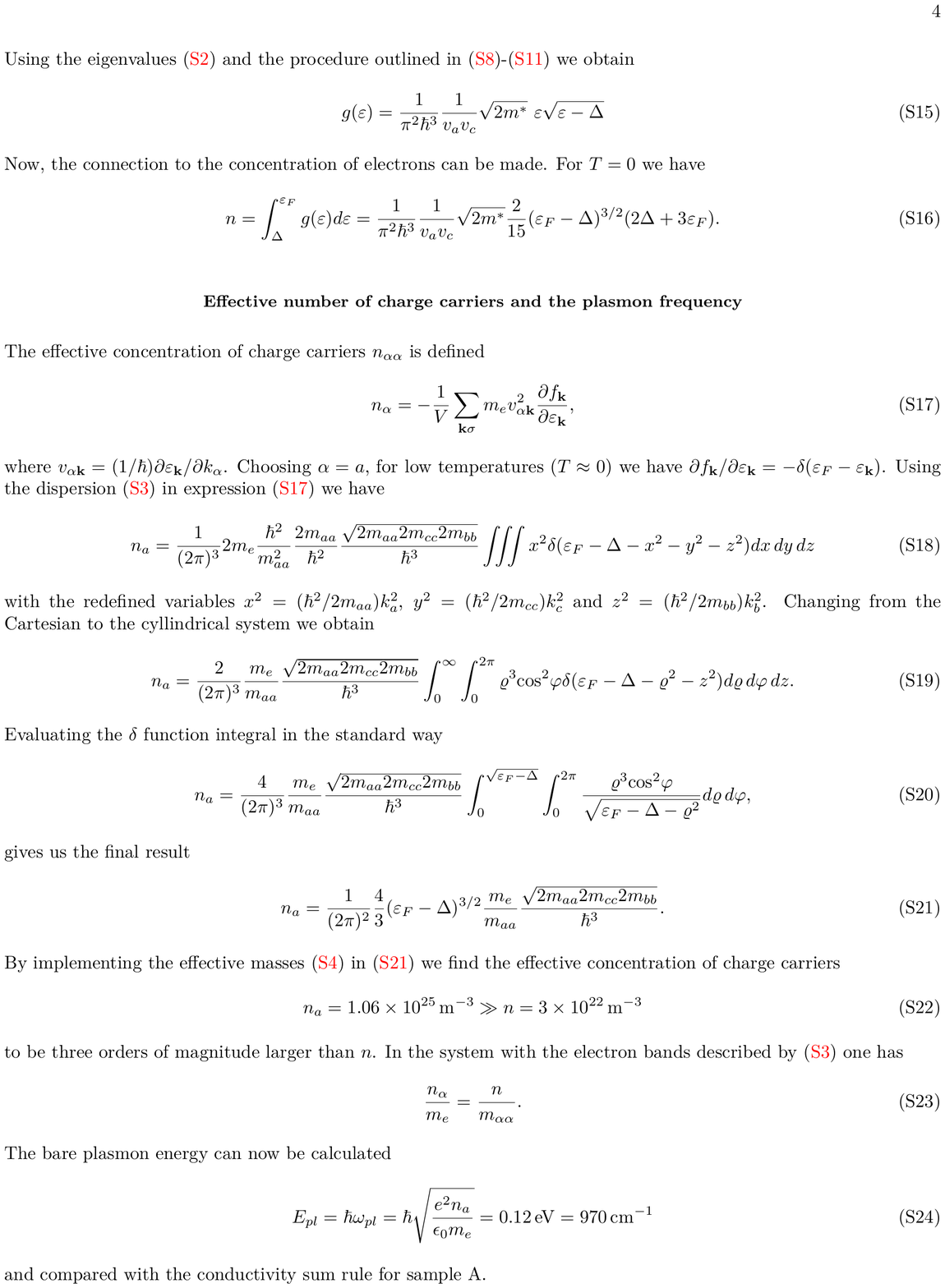} \\
  \ \\
}

\newpage
\vspace*{-2.0cm}
\hspace*{-2.5cm} {
  \centering
  \includegraphics[width=1.2\textwidth]{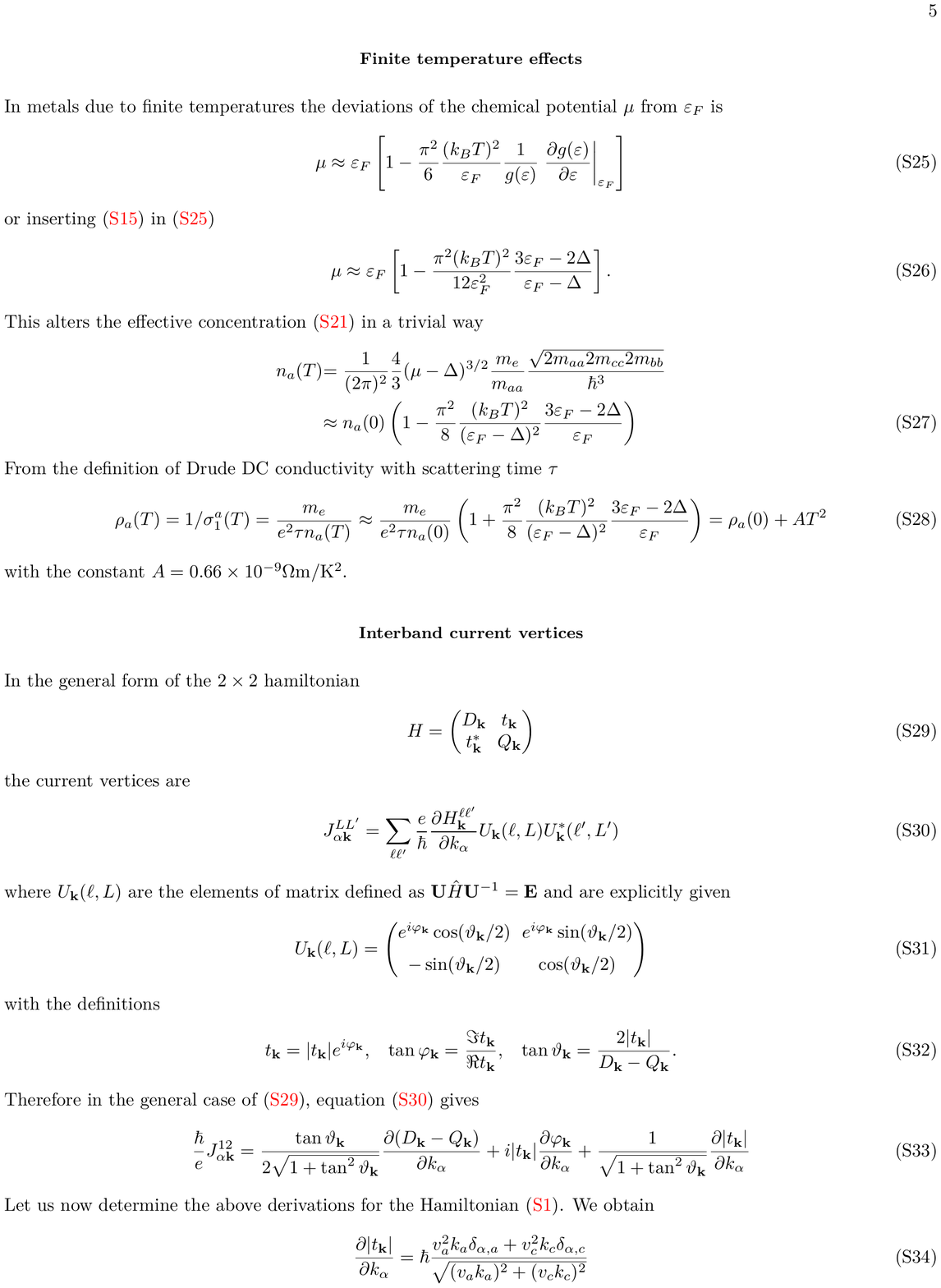} \\
  \ \\
}

\newpage
\vspace*{-2.0cm}
\hspace*{-2.5cm} {
  \centering
  \includegraphics[width=1.2\textwidth]{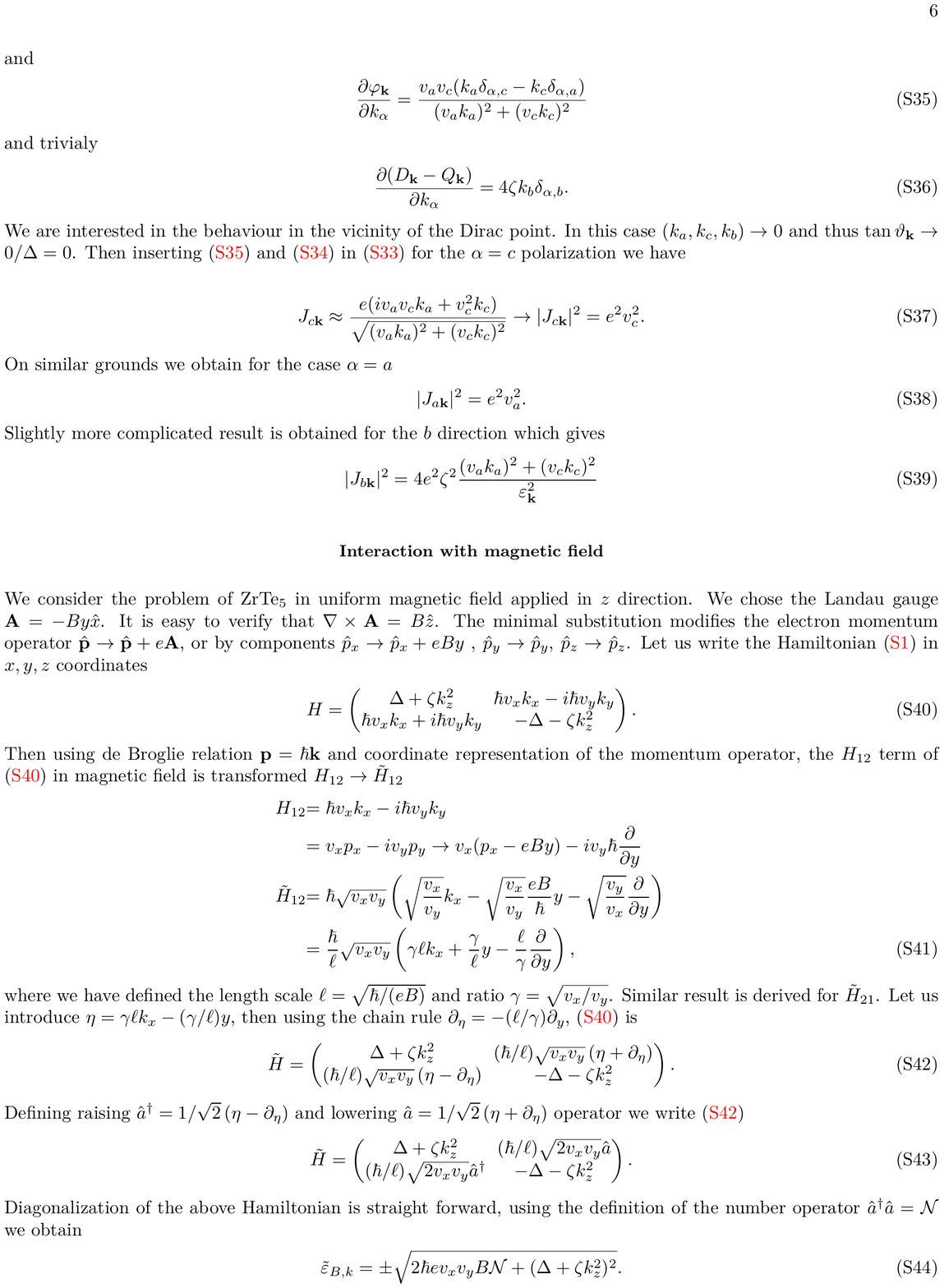} \\
  \ \\
}

\newpage
\vspace*{-2.0cm}
\hspace*{-2.5cm} {
  \centering
  \includegraphics[width=1.2\textwidth]{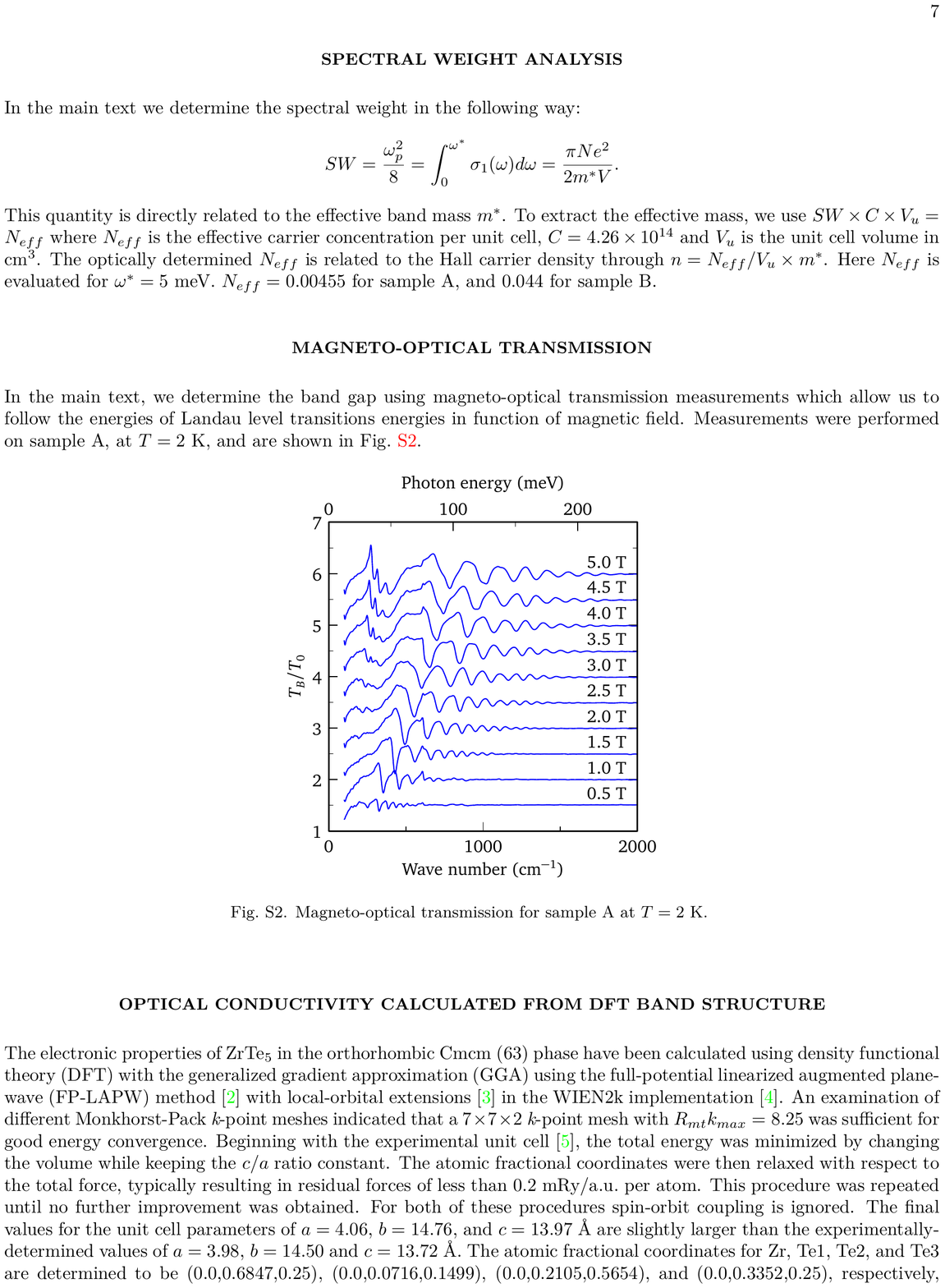} \\
  \ \\
}

\newpage
\vspace*{-2.0cm}
\hspace*{-2.5cm} {
  \centering
  \includegraphics[width=1.2\textwidth]{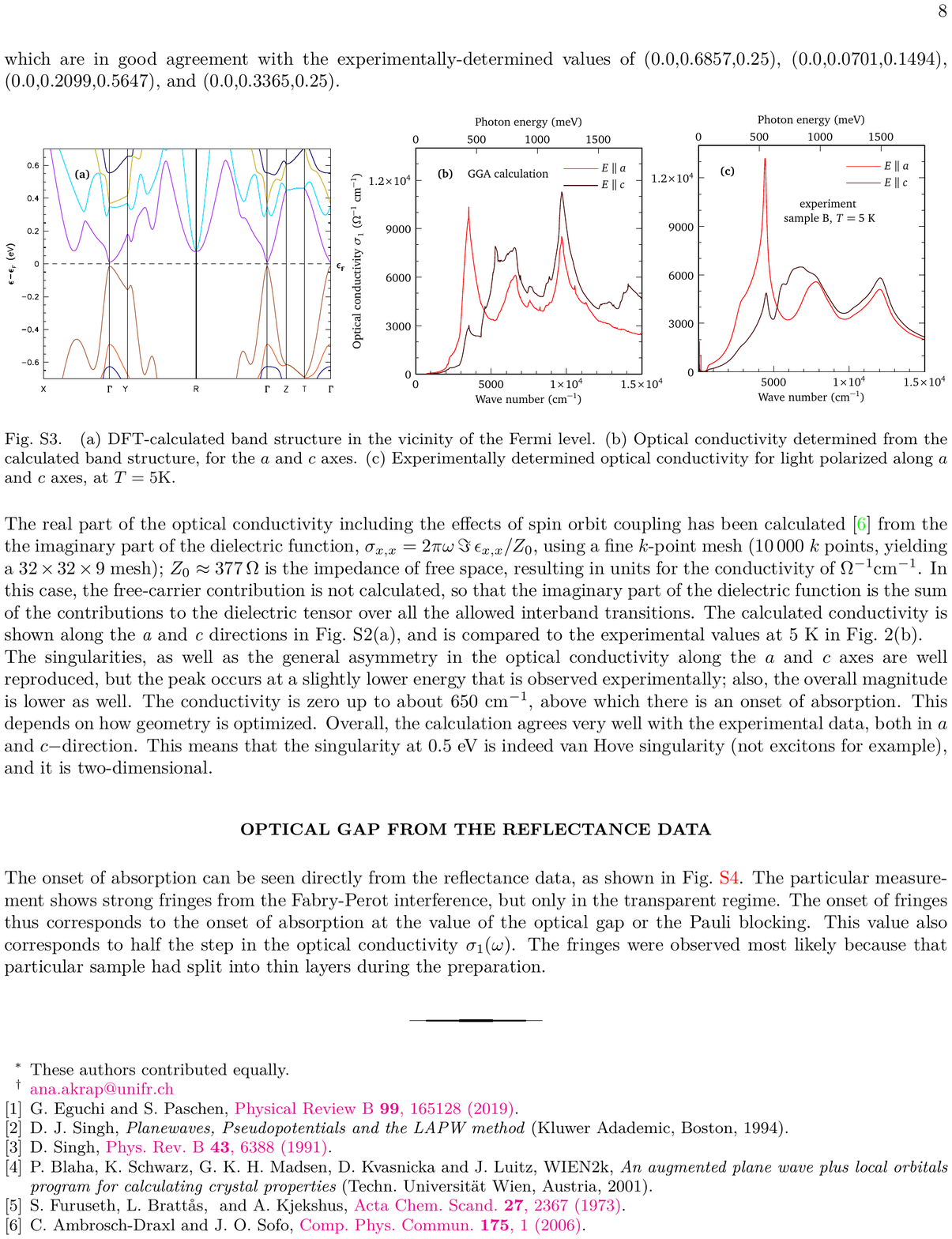} \\
  \ \\
}

\newpage
\vspace*{-2.0cm}
\hspace*{-2.5cm} {
  \centering
  \includegraphics[width=1.2\textwidth]{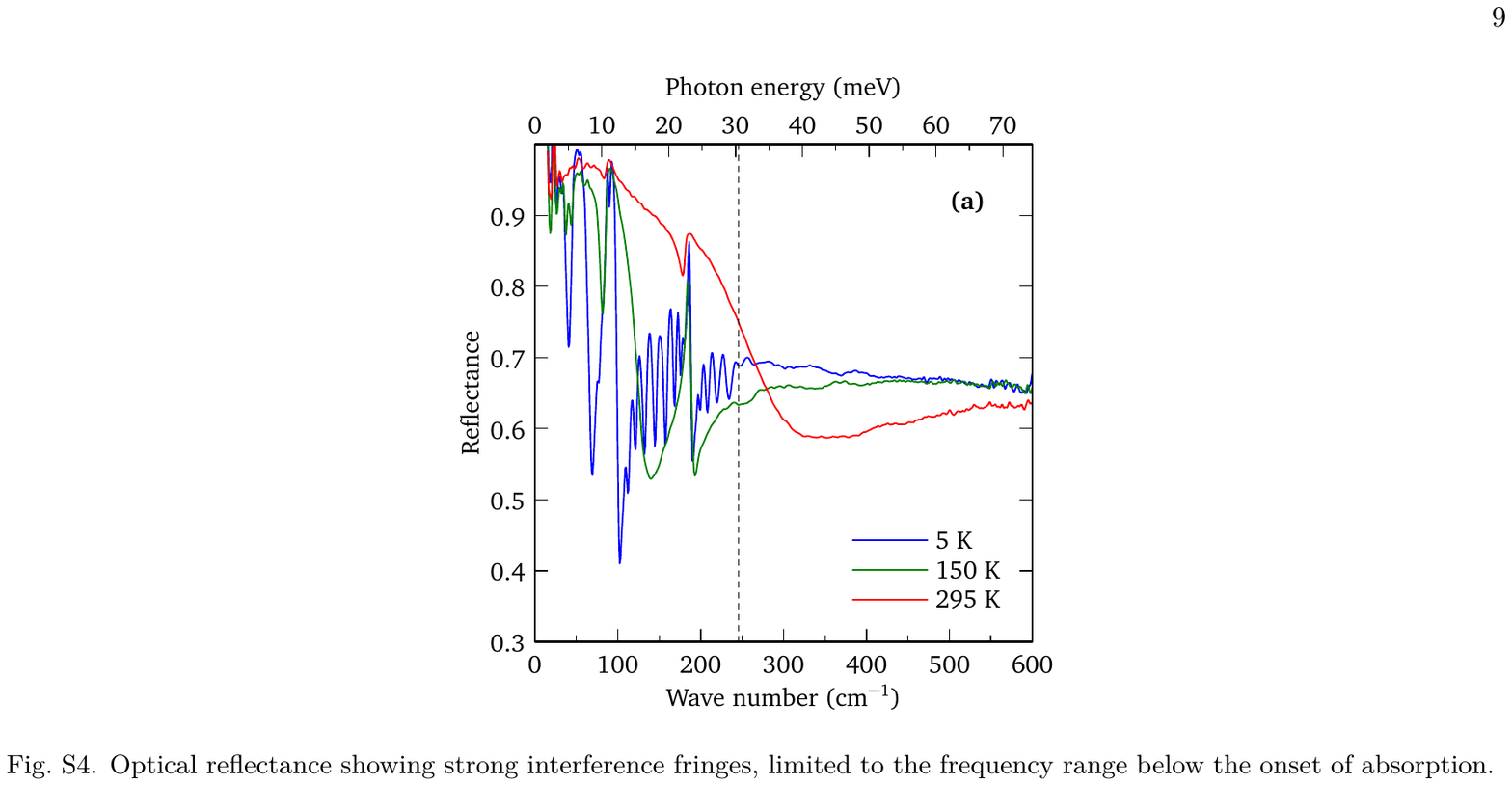} \\
  \ \\
}

\end{document}